\documentclass[twocolumn,showpacs,floatfix,prb,10pt]{revtex4-1}
\usepackage[english]{babel}
\usepackage{amsmath,amssymb}
\usepackage{times}
\usepackage{epsfig}
\usepackage[colorlinks,linkcolor=blue,citecolor=blue]{hyperref}
\usepackage{float}
\newcommand{\wt}{\widetilde}
\begin{document}
\title{Effective realization of random magnetic fields in compounds with large single--ion anisotropy}
\author{J. Herbrych$^{1,2}$}
\author{J. Kokalj$^{3,4}$}
\affiliation{$^1$Department of Physics and Astronomy, The University of Tennessee, Knoxville, Tennessee 37996, USA}
\affiliation{$^2$Materials Science and Technology Division, Oak Ridge National Laboratory, Oak Ridge, Tennessee 37831, USA}
\affiliation{$^3$J. Stefan Institute, SI-1000 Ljubljana, Slovenia}
\affiliation{$^4$Faculty of Civil and Geodetic Engineering, University of Ljubljana, SI-1000 Ljubljana, Slovenia}
\begin{abstract}
We show that spin $S=1$ system with large and random single--ion anisotropy can be at low energies mapped to a $S=1/2$ system with random magnetic fields. This is for example realized in \mbox{Ni(Cl$_{1-x}$Br$_{x}$)$_2$-4SC(NH$_2$)$_2$} compound (DTNX) and therefore it represents a long sought realization of random local (on-site) magnetic fields in antiferromagnetic systems. We support the mapping by numerical study of $S=1$ and effective $S=1/2$ anisotropic Heisenberg chains and find excellent agreement for static quantities and also for the spin conductivity. Such systems can therefore be used to study the effects of local random magnetic fields on transport properties.
\end{abstract}
\date{\today}
\maketitle

\section{Introduction}

In the interacting many--body systems weak disorder usually acts as a source of scattering and leads to a broadening of a Drude peak and increased resistivity. On the other hand, the effect of strong disorder, e.g., large local random magnetic fields, are expected to lead to conceptually novel and more exotic behavior. Examples of these include Bose glass \cite{Yu2012,Ristivojevic2014} (interacting bosons undergoing a phase transition between a superfluid and a localized phase), subdiffusive dynamics (i.e., optical conductivity showing the anomalous power law $\sigma(\omega\to0)\simeq \omega^\alpha$ with $\alpha < 1$), or even a many--body localized phase. The latter, is an interacting analog of Anderson localization \cite{Anderson1958} and the properties of a system close to, at, or in such a phase are a focus of many recent theoretical studies. \cite{Berkelbach2010, Gopalakrishnan2015, Steinigeweg2015,Pal2010,BarLev2015,Luitz2015,Agarwal2015, Barisic2016, Monthus2010, Bera2015, Torres2015, Johri2015, Serbyn2013, Potter2016,Vosk2015,Medvedyeva2016,Kozarzewski2016,Prelovsek2016}

On the other hand, experimental studies of such phenomena are surprisingly rare, mainly due to the lack of real world realizations of strong enough disorder. Recently a few studies of cold atoms on optical lattices \cite{Schreiber2015,Bordia2015,Choi2016} and a study of short ion chains \cite{Smith2015} were preformed. On the other hand, in real materials the disorder is usually introduced by doping, which, e.g., in spin systems locally alters the exchange interactions making the system random \cite{Shiroka2011,Herbrych2013}. Similar off-diagonal disorder is realized in dipolar ferromagnetic Ising compounds \cite{Schechter2006,Silevitch2007,Schechter2008,Wen2010}, e.g., \mbox{LiHo$_x$Y$_{1-x}$F$_4$}, which within a perturbation theory around the ferromagnetic state \cite{Schechter2006} leads to the random magnetic field of the order of exchange interaction. However, to induce and study the effects of strong disorder a systems with stronger local disorder are preferred and needed, e.g., a system with large local random magnetic fields. This is reflected also in a disproportionate large number of theoretical studies based on a $S=1/2$ antiferromagnetic Heisenberg model with random magnetic fields (equivalent to interacting spinless fermions with random on--site energy).

With this work we show that $S=1$ antiferromagnetic Heisenberg model (AHM) with large single--ion anisotropy $D$ realizes an effective low--energy Hamiltonian with locally random magnetic fields when subjected to doping. Such a setup is for example realized in \mbox{Ni(Cl$_{1-x}$Br$_{x}$)$_2$-4SC(NH$_2$)$_2$} compound (DTNX) and therefore it represents a long sought realization of random local (on-site) magnetic fields in antiferromagnetic systems. In particular, we show that in the large $D\gg J$ limit ($J$ is the exchange coupling) the $S=1$ model maps to effective $S=1/2$ model in a magnetic field. I.e., with $D$ being the largest energy scale we can discuss the behavior of the model in local $S=1$ basis: $|-1\rangle$, $|0\rangle$ and $|1\rangle$. For large magnetic fields ($h\gtrsim D$) the states $|0\rangle$ and $|-1\rangle$ are low--lying while the state $|1\rangle$ is by about $2h$ or $2D$ higher in energy. This state can be projected out, while the two low--lying states can be regarded as two states of $S=1/2$, i.e., $|\downarrow\rangle$ and $|\uparrow\rangle$. More importantly, effective model exhibit random on-site magnetic fields. The latter are coming primarily from quenched disorder of single-ion anisotropy.

The paper is organized as follows: in Sec.~\ref{model} we present the $S=1$ model and derivation of the effective $S=1/2$ Hamiltonian. Section~\ref{testing} is devoted to the numerical tests of the mapping for various parameters. Main result, i.e., comparison of the dynamical optical conductivity for $S=1$ and $S=1/2$ system is presented in Sec.~\ref{conductivity}. Finally, conclusions together with the discussion on experimental realization of quenched randomness in the antiferromagnetic $S=1$ compound is given in Sec.~\ref{conclusion}.

\section{Effective low--energy Hamiltonian and random magnetic fields}
\label{model}

Let us start with one--dimensional (1D) $S=1$ AHM
\begin{equation}
H=\sum_{i=1}^{L}\left[J_{i}\mathbf{S}_{i}\cdot\mathbf{S}_{i+1}+D_{i}(S^{z}_{i})^{2}
+hS^{z}_{i}\right]\,,
\label{s1ham}
\end{equation}
with quenched disorder in both exchange coupling $J_i$ and single--ion anisotropy $D_i$, both depending on site index $i$. $\mathbf{S}_{i}=(S^{x}_{i},S^{y}_{i},S^{z}_{i})$ are spin $S=1$ operators at site $i$ and $h$ is the magnetic field. We consider $J_{i}$ and $D_{i}$ as uncorrelated and uniformly distributed in intervals $J-\delta J <J_i< J+\delta J$ (with average $\overline{J}=J$) and $D-\delta D< D_i<D+\delta D$ (with average $\overline{D}=D$). In the reminder of this work we will denote randomness with $\delta R=(\delta J, \delta D)$. Furthermore, we fix the anisotropy $D=4J$ (relevant for DTNX compound \cite{Psaroudaki2012}), and use $J=1$ as energy units, together with $k_{\text{B}}=\hbar=1$.

\begin{figure}[!ht]
\includegraphics[width=1.0\columnwidth]{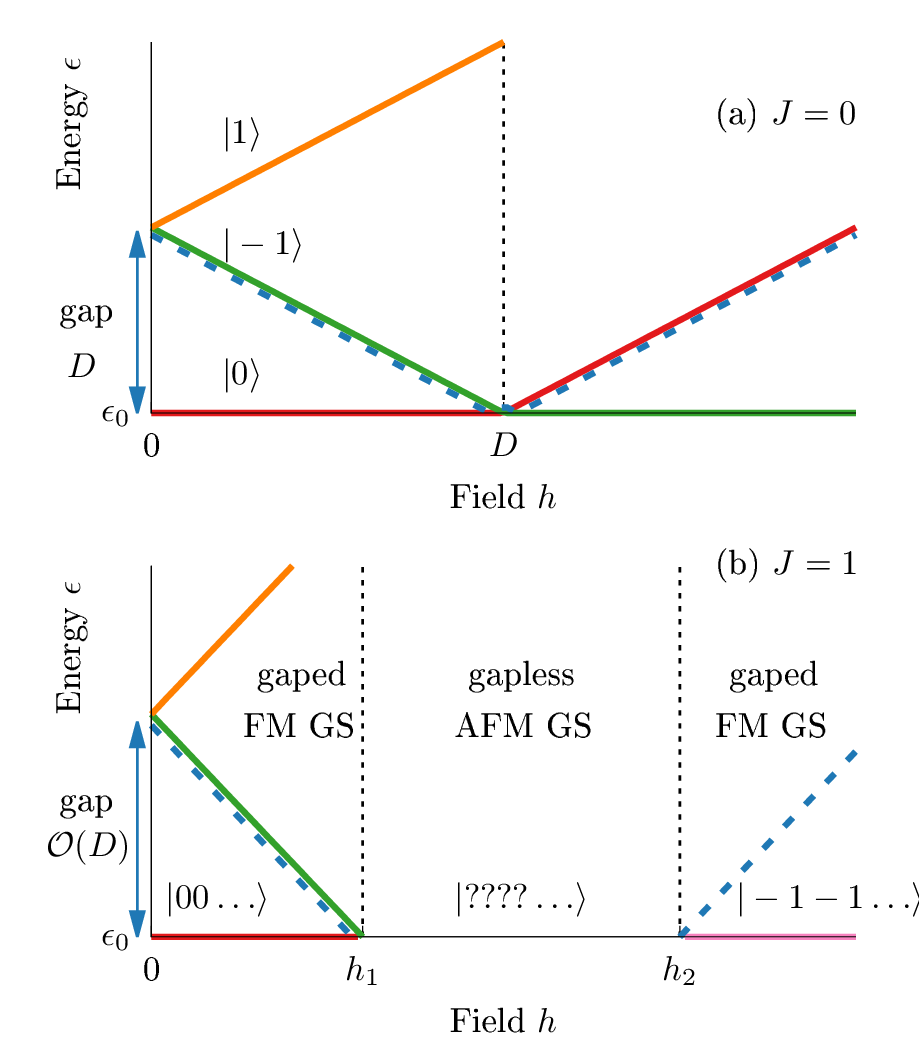}
\caption{(Color online) Sketch of generic $T=0$ magnetic field $h$ phase diagram of (a) $J=0$ and (b) $J\ne0$ system with large single--ion anisotropy $D$. Dashed lines represent closing and opening of energy gap. $\epsilon_0$ denotes the ground--state energy, FM GS-ferromagnetic ground--state, AFM GS-antiferromagnetic ground--state.}
\label{fig_S0}
\end{figure}

The simplest, crude way to justify mapping to the effective model is to consider $J=0$, i.e., $H=\hat{D}$. In such a single--particle picture, the $h=0$ ground--state (GS) is the product of states $|0\rangle$ with degenerated $|-1\rangle$ and $|1\rangle$ excitations separated by $D$ [see Fig.~\ref{fig_S0}(a)]. Next, finite magnetic field $h$ (Zeeman term) splits $|-1\rangle$ and $|1\rangle$. At $h=D$ the $|-1\rangle$ becomes degenerated with $|0\rangle$. Above $h=D$ the GS becomes product of $|-1\rangle$ states. It is obvious that at $h\gtrsim D$ the low--energies can be described only by two states per site, $|-1\rangle$ and $|0\rangle$.

In panel (b) of Fig.~\ref{fig_S0} we sketch the phase diagram of full $S=1$ Hamiltonian with finite exchange interaction $J\ne0$ and large single--ion anisotropy $D$. Critical fields can be calculated with $1/D$ expansion \cite{Papanicolaou1997,Psaroudaki2012}, i.e., $h_{1}=D-2J+J^2/D+J^3/(2 D^2)$ and $h_{2}=D+4J$ which for our choice of anisotropy yields $h_{1}/J\simeq2.28$ and $h_{2}/J=8$.

Let us now describe the mapping of the full $S=1$ model. Our aim is to integrate out the higher energy states $|1\rangle$ or to make the Hamiltonian block diagonal in the subspaces of fixed number of spins in $|1\rangle$. This is similar to strong coupling approach as introduced in Ref. \onlinecite{Eskes1994}, where the Hubbard Hamiltonian was made block diagonal in the subspaces of fixed number of doubly occupied sites. The used effective $S=1/2$ model $\wt{H}$ will be the lowest energy block, without spins in state $|1\rangle$. 

Let us start with the unitary transformation of the form
\begin{equation}
H=\mathrm{e}^{-S}H^{\prime}\mathrm{e}^{S}=H^{\prime}+[S,H^{\prime}]
+\frac{1}{2}[S,[S,H^{\prime}]]]+\cdots\,,
\label{S_reftra}
\end{equation}
such that the lowest order terms in $J/D$ of the above expansion will not change the number of $|1\rangle\,$ states. Such a requirement is equivalent to making the number of $|1\rangle\,$ states a good quantum number. It is convenient to first rewrite the $S=1$ Hamiltonian \eqref{s1ham} by using the notation and the approach of Ref.~\onlinecite{Eskes1994}.
\begin{equation}
H = \sum_{i=1}^{L}\left(T^{\pm}_{i}+T^{zz}_{i}+\hat{D}_{i}\right)\,,
\label{S_s1ham}
\end{equation}
where
\begin{eqnarray}
T^{\pm}_i &=& \frac{J_i}{2}\left(S^{+}_{i}S^{-}_{i+1}+S^{-}_{i}S^{+}_{i+1}\right)\,,
\qquad T^{\pm}=\sum_i^L T^{\pm}_{i}\,,\nonumber\\
T^{zz}_i &=& J_iS^{z}_{i}S^{z}_{i+1}\,,\qquad T^{zz}=\sum_i^L T^{zz}_{i}\,,\nonumber\\
\hat{D}_i &=& D_{i}(S^{z}_{i})^{2}\,,\qquad \hat{D}=\sum_i^L \hat{D}_{i}.\nonumber
\end{eqnarray}
It is obvious that $T^{zz}_i$ and $\hat{D}_i$ in \eqref{S_s1ham} do not change the number of $|1\rangle\,$  states. On the other hand, 
\begin{equation}
T^{\pm1}_{i} = T^{-1}_{i}+T^{0}_{i}+T^{+1}_{i}\,,\nonumber
\end{equation}
can increase ($T^{+1}_{i}$), decrease ($T^{-1}_{i}$), or leave unchanged ($T^{0}_{i}$) the total number of $|1\rangle$ states. Operators $T^{-1,0,+1}_{i}$ can be defined with help of projection operators in the ($|1\rangle$, $|0\rangle$, $|-1\rangle$) basis
\begin{equation}
n^{1}_{i}=
\begin{pmatrix}
1 & 0 & 0\\ 0 & 0 & 0\\ 0 & 0 & 0\\
\end{pmatrix}\,,\quad
n^{0}_{i}=
\begin{pmatrix}
0 & 0 & 0\\ 0 & 1 & 0\\ 0 & 0 & 0\\
\end{pmatrix}\,,\quad
n^{-1}_{i}=
\begin{pmatrix}
0 & 0 & 0\\ 0 & 0 & 0\\ 0 & 0 & 1\\
\end{pmatrix}\,.\nonumber
\end{equation}
Obviously $n^{1}_{i}+n^{0}_{i}+n^{-1}_{i}=\mathbb{I}_i$. Explicit form is given by
\begin{eqnarray}
T^{-1}_{i} = \frac{J_i}{2}(n^{0}_{i}S^{+}_{i}S^{-}_{i+1}n^{1}_{i+1}
&+& n^{0}_{i}S^{-}_{i}S^{+}_{i+1}n^{-1}_{i+1})\,,\nonumber\\
\nonumber\\
T^{0}_{i} = \frac{J_i}{2}(n^{0}_{i}S^{+}_{i}S^{-}_{i+1}n^{0}_{i+1}
&+& n^{-1}_{i}S^{-}_{i}S^{+}_{i+1}n^{-1}_{i+1})\nonumber\\
+ \frac{J_i}{2}(n^{1}_{i}S^{+}_{i}S^{-}_{i+1}n^{1}_{i+1}
&+& n^{0}_{i}S^{-}_{i}S^{+}_{i+1}n^{0}_{i+1})\,,\nonumber\\
\nonumber\\
T^{+1}_{i} = \frac{J_i}{2}(n^{1}_{i}S^{+}_{i}S^{-}_{i+1}n^{0}_{i+1}
&+& n^{-1}_{i}S^{-}_{i}S^{+}_{i+1}n^{0}_{i+1})\,.\nonumber
\end{eqnarray}
Next, let us consider $H^{\prime}$ in the same form as \eqref{S_s1ham}
\begin{equation}
H^{\prime} = \sum_{i=1}^{L}
\left[\frac{J_i}{2}\left(S^{+}_{i}S^{-}_{i+1}+S^{-}_{i}S^{+}_{i+1}\right)
+J_{i}S^{z}_{i}S^{z}_{i+1}+D_{i}(S^{z}_{i})^{2}\right]\,,\nonumber
\end{equation}
which can be again written as $H^{\prime}=T^{\pm1}+T^{zz}+\hat{D}$. The lowest order of Eq.~\eqref{S_reftra} will conserve the number of $|1\rangle$ states, if $T^{-1}$ and $T^{+1}$ in the first term $H'$ of right hand side of Eq.~\eqref{S_reftra} will cancel with $[S,H']$, namely if
\begin{equation}
[S,H^{\prime}]=-T^{-1}-T^{+1}\,.\nonumber
\end{equation}
Here $T^{\pm1}=\sum_i^L T^{\pm1}_{i}$. Since we are dealing with large--$D$ system, we can rewrite the above equation as $[S,H^{\prime}]=[S_1,\hat{D}]+{\cal O}(J^2/D)$, or
\begin{equation}
[S_{1},\hat{D}]=-T^{-1}-T^{+1}\,,
\end{equation}
where we consider only terms up to $J/D$. One can show that the above equation will be fulfilled by
\begin{equation}
S_{1}=\sum_{i=1}^{L}\frac{-T^{-1}_{i}-T^{+1}_{i}}{2(D_{i}+D_{i+1})}\,.
\end{equation}
Finally, we can write
\begin{eqnarray}
H &=& \mathrm{e}^{-S_{1}}H^{\prime}\mathrm{e}^{S_{1}}=H^{\prime}+[S_{1},H^{\prime}]
+\frac{1}{2}[S_{1},[S_{1},H^{\prime}]]]+\cdots\nonumber\\
&=& H^{\prime} + [S_{1},\hat{D}]+\cdots= T^{0}+T^{zz}+\hat{D}+{\cal O}(J^2/D)\,,\nonumber
\end{eqnarray}
where $T^{0}=\sum_i^L T^{0}_{i}$. Let us now write the explicit form of the above ($1/D$ approximate) Hamiltonian
\begin{eqnarray}
T^{0}_{i} &+& T^{zz}_{i}+\hat{D}_{i} = \wt{H}+\wt{H}_{1}\nonumber\\
\wt{H} &=& \frac{J_i}{2}n^{0}_{i}S^{+}_{i}S^{-}_{i+1}n^{0}_{i+1}
+ \frac{J_i}{2}n^{-1}_{i}S^{-}_{i}S^{+}_{i+1}n^{-1}_{i+1}\nonumber\\
&+& J_{i}n^{-1}_{i}S^{z}_{i}S^{z}_{i+1}n^{-1}_{i+1}
+ D_{i}(S^{z}_{i})^{2}n^{-1}_{i}\nonumber\\
&+& J_{i}n^{0}_{i}S^{z}_{i}S^{z}_{i+1}n^{0}_{i+1}
+ D_{i}(S^{z}_{i})^{2}n^{0}_{i}\,,\label{hths1}\\
\wt{H}_{1} &=& \frac{J_i}{2}n^{1}_{i}S^{+}_{i}S^{-}_{i+1}n^{1}_{i+1}
+ \frac{J_i}{2}n^{0}_{i}S^{-}_{i}S^{+}_{i+1}n^{0}_{i+1}\nonumber\\
&+& J_{i}n^{-1}_{i}S^{z}_{i}S^{z}_{i+1}n^{1}_{i+1}
+ J_{i}n^{1}_{i}S^{z}_{i}S^{z}_{i+1}n^{-1}_{i+1}\nonumber\\
&+& J_{i}n^{1}_{i}S^{z}_{i}S^{z}_{i+1}n^{1}_{i+1} + D_{i}(S^{z}_{i})^{2}n^{1}_{i+1}\,.
\end{eqnarray}
We see that the above Hamiltonian does not mix the states with different number of spins in $|1\rangle$, or that $[\wt{H},\sum_i n_i^1]=0$ and also $[\wt{H}_1,\sum_i n_i^1]=0$. At this point the mixing terms are of higher order ($J^2/D$). It is further clear that $\wt{H}$ is nonzero only for states with no spins in state $|1\rangle$. 

In the presence of finite magnetic field $h$ the lowest energy from the $\wt{H}_{1}$ sub--system will be at least ${\cal O}(D+h)$ for $h\sim J$ (and even ${\cal O}(D+2h)$ for $h\gg J$) higher than the lowest states of $\wt{H}$. As a consequence, within such a $h$ region the low--energy (i.e., low temperature) properties of the system can be described solely by $\wt{H}$ block. Since $\wt{H}$ is spanned by $|-1\rangle$ and $|0\rangle$ we can omit the projection operators, and use transformation $S^z=\wt{S}^z+1/2\,,S^{\pm}=\sqrt{2}\wt{S}^{\pm}$. The latter maps $|-1\rangle\to|\downarrow\rangle$ and $|0\rangle\to|\uparrow\rangle$. Finally we can write $\wt{H}$ (together with Zeeman term $hS^z_i\to h \wt{S}^z_i$) as the anisotropic $S=1/2$ Heisenberg model
\begin{equation}
\wt{H}=\sum_{i=1}^{L}\left[\wt{J}_{i}
\!\left(\wt{S}_{i}^{x}\wt{S}_{i+1}^{x}+\wt{S}_{i}^{y}\wt{S}_{i+1}^{y} 
+\Delta\wt{S}_{i}^{z}\wt{S}_{i+1}^{z}\!\right)+\wt{h}_{i}\wt{S}_{i}^{z}\right]\,.
\label{s12ham}
\end{equation}
Here $\wt{S}^{\alpha}_{i}$ with $\alpha=x,y,z$ are spin $S=1/2$ operators at site $i$, $\wt{J}_i=2J_i$, $\Delta=0.5$ is an exchange anisotropy, and \mbox{$\wt{h}_{i}=h-D_{i}-(J_{i}+J_{i-1})/2$}. We denote with $\wt{h}$ the average magnetic field in \eqref{s12ham} and the distribution span with $\delta\wt{h}=\delta J+\delta D$. We stress that randomness in $D_i$ and $J_i$ leads to randomness in local magnetic field $\wt{h}_i$ of the effective model. For a case with randomness only in $D_i$, one would have random magnetic field $S=1/2$ Heisenberg model. Note also that the average effective magnetic filed $\wt{h}$ is decreased from $h$ by $D+J$ and vanishes for $h=D+J$. Furthermore, $S=1/2$ model predicts the same second critical filed, $\wt{h}_2=D+4J$, while for the first one gives correctly the first order in terms of $J/D$, i.e., $\wt{h}_1=D-2J$.

\section{Test of the mapping}
\label{testing}

In the following we compare several static and dynamic quantities obtained with the full $S=1$ model \eqref{s1ham} with those obtained with the effective $S={1/2}$ model \eqref{s12ham} in order to support the mapping and determine its regime of applicability. Most of the quantities are calculated with Lanczos for ground state or finite--temperature Lanczos method (FTLM) \cite{prelovsek2013} on finite chains with $L=14$ sites and by using $\sim 20$ initial Lanczos vectors and $M=400$ Lanczos steps. In addition we support Lanczos results also with results from transfer matrix renormalization group (TMRG) \cite{Shibata1997,Wang1997,Psaroudaki2014} for $L=\infty$ (pure system only) and density matrix renormalization group (DMRG) \cite{Schollwock2005} with $L=800$.

\begin{figure}[!ht]
\includegraphics[width=1.0\columnwidth]{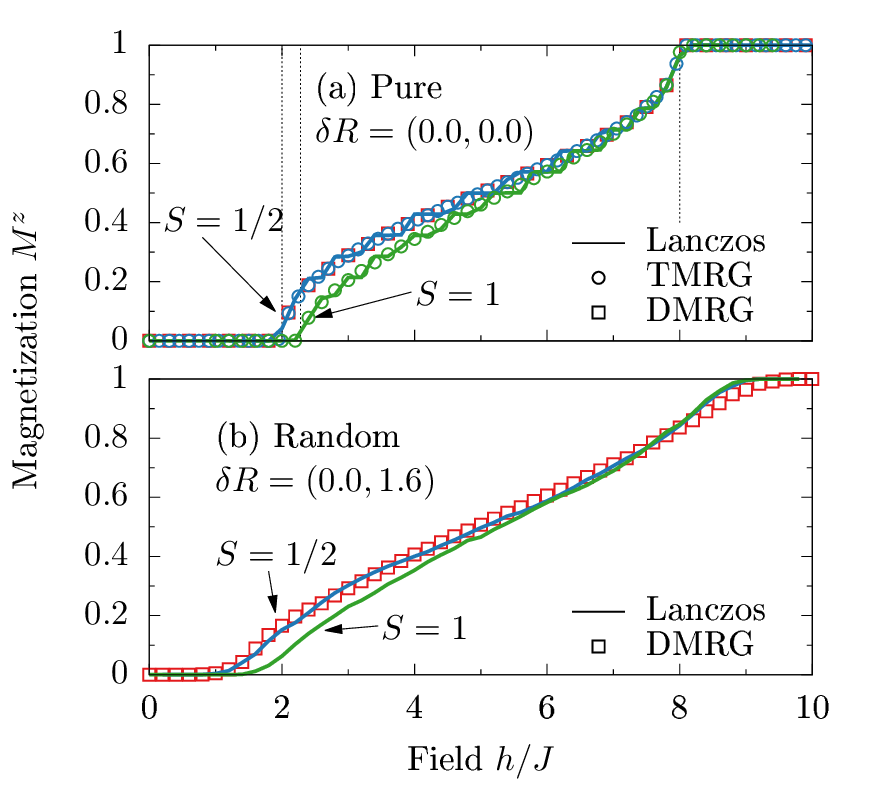}
\caption{(Color online) $T=0$ magnetization $M^z$ vs. magnetic field $h$ of the effective $S=1/2$ model shows agreement with the full $S=1$ model. Panel (a) shows results for pure case $\delta R=(0.0,0.0)$, while panel (b) for random system with $\delta R=(0.0,1.6)$. $T=0$ and $D/J=4$. Vertical lines represent critical fields. Results are obtained with Lanczos method ($L=14$, $N_{\text{R}}=200$ realizations of random system), TMRG ($L=\infty$, $N_{\text{R}}=1$ taken from Ref.~\onlinecite{Psaroudaki2014}) and DMRG ($L=800$, $N_{\text{R}}=1$ random realization, 200 basis states kept).}
\label{fig_1}
\end{figure}

In Fig.~\ref{fig_1} we show $h$ dependence of magnetization $M^z=- \sum_{i=1}^L \langle S_i^z\rangle/L$ for pure $\delta R=(0.0,0.0)$ and random $\delta R=(0,1.6)$ cases at $T=0$. $\langle\ldots\rangle$ denotes the thermodynamic average at temperature $T$ and average over $N_r$ configurations of $J_i$ and $D_i$. It is worth noting that, although hereafter we present results only in the ergodic (thermal) phase, in the localized phase the system is not ergodic and therefore does not thermalize. As a consequence, in such a phase the used Boltzmann thermal average and the notion of temperature is invalid, and one should instead explore the behavior of a representative state, which, e.g., depends on the preparation protocol and has a characteristic energy density. 

We first note that the comparison of Lanczos results with the TMRG and DMRG results is satisfactory, giving the support to the Lanczos approach.  Fig.~\ref{fig_1}(a) shows results for a pure $\delta R=(0.0,0.0)$ system, for which $M^z$ stays zero up to the first critical field $h_1$. This is due to the gapped magnon excitations for large $D$ \cite{Papanicolaou1997,Psaroudaki2012}. $h_{1}/J\simeq2.28$ for $S=1$ model while it is slightly lower for effective $S=1/2$ model due to higher order corrections of the $1/D$ expansion \cite{Psaroudaki2012}. With increasing $h$ both models give very similar increase of $M^z$ and at the higher critical field $h_2/J=8$ show perfect agreement. At $h_2$ one enters into a fully polarized ferromagnetic state. In Fig.~\ref{fig_1}(b) similar results are shown for random case with $\delta R=(0,1.6)$. It is clear that sharp features at $h_1$ and $h_2$ shown in Fig.~\ref{fig_1}(a) for pure case are now broadened due to randomness. More importantly, results for effective $S=1/2$ model agree qualitatively and for larger $h$ also quantitatively with the results for full $S=1$ model. This gives strong support for the description of low energy physics of the $S=1$ model \eqref{s1ham} with the $S=1/2$ model \eqref{s12ham} in a wide range of $h$.

Note that due to spin--inversion symmetry, the $S=1/2$ results are symmetric with respect to $h=D+J$ ($\wt{h}=0$), e.g., $h/J=5$ for case shown in Fig.~\ref{fig_1}, while no such symmetry is present for $S=1$ model. Difference is again due to higher order terms in $1/D$ expansion. We also note that our results qualitatively agree with experimental observations on doped DTNX \cite{Yu2012}. In particular, increasing disorder (i) reduces (increases) first (second) critical field $h_1$ ($h_2$), and (ii) increases the critical exponent $\phi$ with which magnetization approaches critical fields $|h-h_{1,2}|^\phi$.

\begin{figure}[!ht]
\includegraphics[width=1.0\columnwidth]{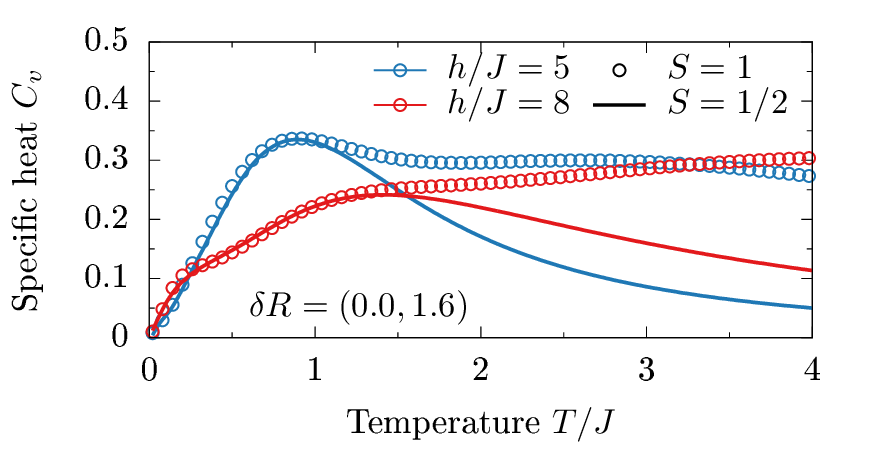}
\caption{(Color online) Specific heat $C_v$ of a $S=1$ system is at $T\lesssim J$ very well captured by effective $S=1/2$ model. Results are calculated for $h/J=5$ and $8$ ($\wt{h}=0$ and $3$), $\delta R=(0.0,1.6)$ and are obtained with Lanczos method ($L=14$, $N_{\text{R}}=200$ random realizations).}
\label{fig_2}
\end{figure}

Above we compared results for $T=0$ where the effective low--energy Hamiltonian is expected to work well. In the following we focus on finite $T$ and show that effective $S=1/2$ model gives satisfactory description also for finite temperatures. In Fig.~\ref{fig_2} we show comparison of static quantity, namely specific heat 
\begin{equation}
C_v=\frac{\langle H^2\rangle-\langle H\rangle^2}{T^2L}\,,
\end{equation}
where
\begin{equation}
\langle H^2\rangle=\sum_{n} p_{n}\epsilon_{n}^{2}\,,
\qquad
\langle H\rangle=\sum_{n}p_{n} \epsilon_{n}\,.
\nonumber
\end{equation}
$p_n=\exp(-\beta\epsilon_n)/Z$ denotes the Boltzmann factor for the eigenstate with energy $\epsilon_n$. For presented $h/J=5$ and $8$ ($\wt{h}/J=0$ and $3$) and $\delta R=(0,1.6)$ and find a very good agreement between $S=1/2$ and $S=1$ models up to $T \sim J$. Fig.~\ref{fig_2} also nicely demonstrates how effective $S=1/2$ model captures only the low lying excitations related to local states $|-1\rangle$ and $|0\rangle$, while it misses the higher energy ones related to $|1\rangle$.

\section{Spin conductivity}
\label{conductivity}

\begin{figure}[!ht]
\includegraphics[width=1.0\columnwidth]{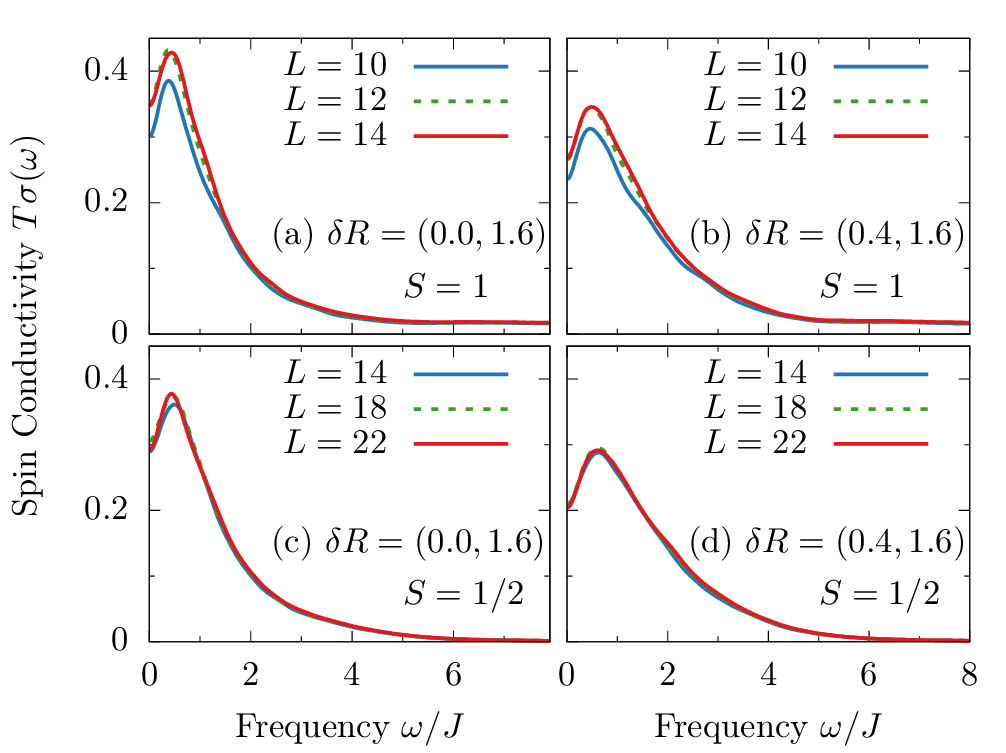}
\caption{(Color online) Finite--size dependence of spin conductivity $T\sigma(\omega)$ for $T/J=1$, $h/J=5$ and $S=1$ ($L=10,12,14$) and $S=1/2$ ($L=14,18,22)$ systems. Panels (a) and (c) show the results for $\delta R=(0.0,1.6)$, while panels (b) and (d) show the results for $\delta R=(0.4,1.6)$. Results are obtained with the use of FTLM, averaged over $N_{\text{R}}=200$ realizations of random system, and smoothed with $\eta=0.2$.}
\label{fig_S4}
\end{figure}

Disorder is expected to affect most dramatically the transport properties and here we discuss dynamical spin conductivity $\sigma(\omega)$. In the following we show that also $\sigma(\omega)$ of a disordered $S=1$ model behaves as a $\sigma(\omega)$ of the effective random magnetic field $S=1/2$ model. $\sigma(\omega)$ is given by
\begin{equation}
\sigma(\omega)=\frac{\pi}{L}\frac{1\!-\!\exp(-\beta\omega)}{\omega}
\!\sum_{n,m}p_n |\langle n|j^{z}|m\rangle|^2\delta(\omega\!-\!\epsilon_m\!+\!\epsilon_n)\,,
\label{sgima}
\end{equation}
where $j^{z} = \sum_{i} J_i\left(S^{x}_{i} S^{y}_{i+1} - S^{y}_{i} S^{x}_{i+1}\right)$ is a spin current, $\beta=1/T$. Since our numerical calculations are performed on finite chains, $\sigma(\omega)$ is a sum of weighted $\delta$ functions that need to be smoothed. We used smoothing $\eta=0.2$ which roughly corresponds to energy resolution of our method, i.e., $\Delta\epsilon/M$ where $\Delta \epsilon$ is an energy span. In Fig.~\ref{fig_S4} we present finite-size scaling of spin conductivity for $S=1$ and corresponding $S=1/2$ model. As evident, results for two largest considered $L$ are almost indistinguishable. 

\begin{figure}[!ht]
\includegraphics[width=1.0\columnwidth]{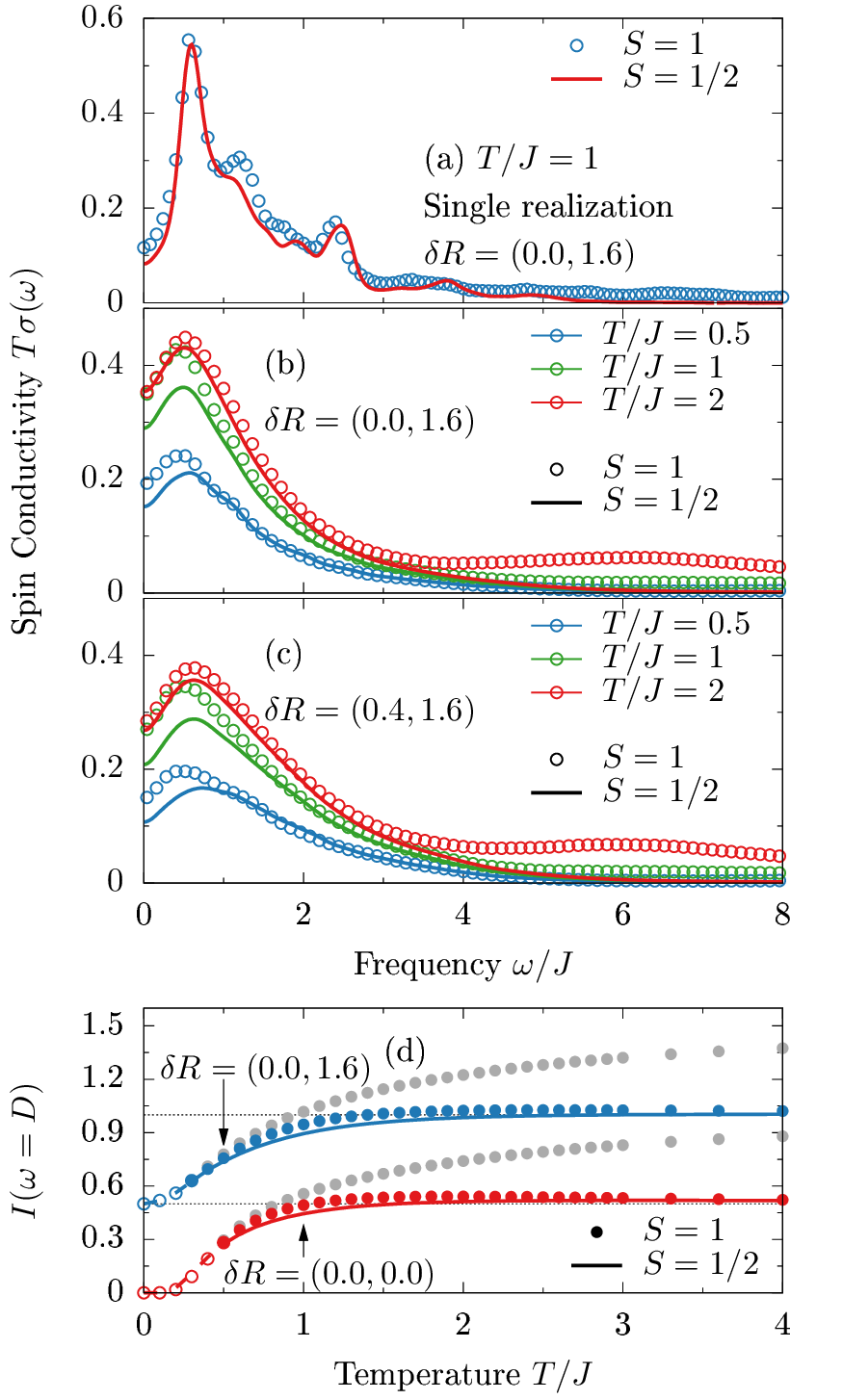}
\caption{(Color online) Effective $S=1/2$ model with random magnetic fields nicely captures the behavior of spin conductivity $\sigma(\omega)$ of the disordered $S=1$ model. Spin conductivities $T\sigma(\omega)$ for $S=1$ and effective $S=1/2$ models and for magnetic field $h/J=5$ ($\wt{h}/J=0)$ is shown for (a) temperature $T/J=1$ and one realization ($N_{\text{R}}=1$) for $\delta R=(0.0,1.6)$. Panels (b) and (c) similarly show $T\sigma(\omega)$ averaged over $N_{\text{R}}=200$ realizations for various temperatures and for $\delta R=(0.0,0.1.6)$ and $\delta R=(0.4,1.6)$. (d) $T$ dependence of integrated spin conductivity $I(\omega=D)$ for $S=1$ and $S=1/2$ system for $h/J=5$, $\delta R=(0.0,0.0)$ and $\delta R=(0.0,1.6)$. The latter is shifted by $+0.5$ for clarity. Grey points depict $I(\omega\to\infty)$ for $S=1$ system. Horizontal dashed line represent exact high--$T$ kinetic energy $T\epsilon_{\text{kin}}=I(\omega\to\infty)$ for $S=1/2$ system. At very low--$T$ open point/dashed lines are used due to possible finite--size effects.}
\label{fig_3}
\end{figure}

In Fig.~\ref{fig_3} we present one of our main results - the comparison of the $\sigma(\omega)$ between the random $S=1$ and the effective $S=1/2$ model for $D/J=4$ and $h/J=5$. We choose such $h$ in order to have an effective $S=1/2$ model with random magnetic fields $\wt{h}_i$ distributed around zero average magnetic field $\wt{h}=0$. In panel (a) we compare $\sigma(\omega)$ for $S=1$ and for effective $S=1/2$ model for one single randomness realization and find very good agreement. This supports the mapping even on the level of small chains, single realization and for transport quantities. In panels (b) and (c) of Fig.~\ref{fig_3} we present conductivity averaged over $N_{\text{R}}=200$ realizations for several $T$ and for $\delta R=(0.0,1.6)$ and $\delta R=(0.4,1.6)$. As expected, the agreement is very good, being qualitative and even quantitative in broad range of $\omega$ (in particular at low $\omega$), $T$ and $\delta R$. This gives strong support that even transport properties of a $S=1$ model can essentially be captured with $S=1/2$ model. It is also clear from comparison of panel (b) and (c) that randomness in $J_i$ or $\wt{J}_i$ has smaller effect on $\sigma(\omega)$ than randomness in $D_i$ or $\wt{h}_i$. Deviations between the two models are expected at higher $T$ and large $\omega$ since the effective $S=1/2$ does not include the higher energy states. This is nicely seen for $\omega>4$ and $T\gtrsim2$ in panels (b) and (c) of Fig.~\ref{fig_3}, were the $S=1/2$ model is missing the high--$\omega$ spectral weight. The agreement for $\omega<4$ even for $T=2$ indicates, that at even such high--$T$ the contribution to Eq.~\eqref{sgima} of higher energy states in small. This is also clearly visible in Fig.~\ref{fig_3}(d) were we present temperature dependence of integrated low-$\omega$ part of spin conductivity $I(\omega)=T/\pi\int_{-\omega}^{\omega}\mathrm{d}\omega^\prime\,\sigma(\omega^\prime)$ for $\omega=D=4J$. Note that for $S=1/2$ system $I(\omega=D)$ exhausts the total sum--rule related to the total kinetic energy of the system $I(\omega=D)\simeq I(\omega=\infty)=T\epsilon_{\text{kin}}$. The latter can be calculated exactly in the high--$T$ limit, i.e., $T\epsilon_{\text{kin}}=\langle j^{z}j^{z}\rangle/L=\wt{J}^2/8$. It is evident that the high--$\omega$ contributions become important for $T\gtrsim J$. The comparison is on the other hand expected to be even better for cases were mapping works better, e.g., for larger $h$ (in particularly close to $h_2$) or for larger $D$. 

\begin{figure}[!ht]
\includegraphics[width=1.0\columnwidth]{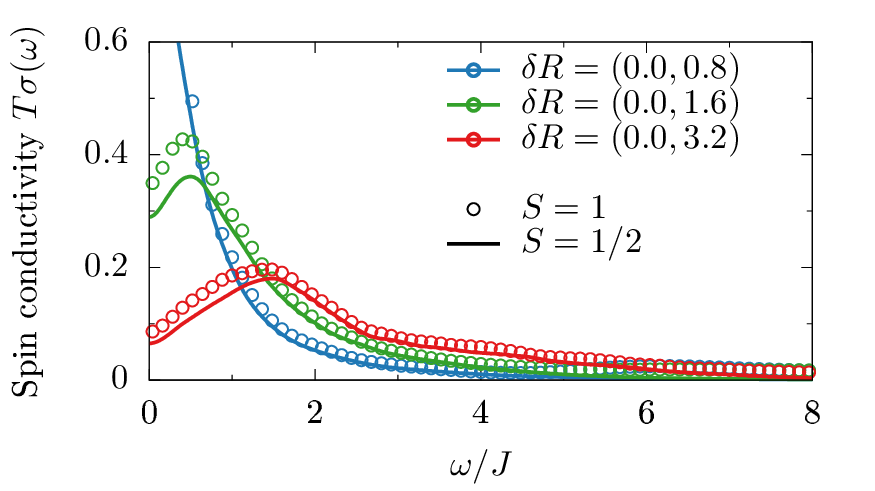}
\caption{(Color online) With increasing randomness both $S=1$ and effective $S=1/2$ model show decreasing low $\omega$ spin conductivity $T\sigma(\omega)$, which is typical for system close to MBL regime. $T/J=1$ and $h/J=5$ ($\wt{h}/J=0)$.}
\label{fig_4}
\end{figure}

Finally, in Fig.~\ref{fig_4}, we present how optical conductivity changes with increasing randomness. Presented results are consistent with decreasing d.c. conductivity with increasing $\delta R$ and thus $\delta\wt{h}$. This is a general behavior of a system with strong and increasing randomness and our results for $\sigma(\omega)$ already compare nicely with some random magnetic field studies \cite{Barisic2010, Steinigeweg2015, Barisic2016}.

\section{Discussion and conclusions}
\label{conclusion}

Let us comment on experimental realization of quenched randomness in antiferromagnetic $S=1$ compound with large, single--ion anisotropy, i.e., \mbox{Ni(Cl$_{1-x}$Br$_{x}$)$_2$-4SC(NH$_2$)$_2$} (DTNX) \cite{Zheludev2013}. The low--energy physics of the clean parent ($\delta J=\delta D=0$) is well studied experimentally and understood theoretically \cite{Zapf2006,Zvyagin2007,Zvyagin2008,Sun2009,Kohama2011,Mukhopadhyay2012,Psaroudaki2012}. Reported values system parameters are $J_{\text{Cl}}\simeq2.2\,\mathrm{K}$, $D_{\text{Cl}}\simeq8.9\,\mathrm{K}$, with out of the chain interaction $J_\perp\simeq0.12\,\mathrm{K}$. Random system is believed to be a mixture of $J_{\text{Cl}}$ and $D_{\text{Cl}}$ with correlated $J_{\text{Br}}$ and $D_{\text{Br}}$ on Br--doped site. Parameters of random Hamiltonian where fitted to reproduce the experiment \cite{Yu2012} and found to be $J_{\text{Br}}/J_{\text{Cl}}\simeq2.35$, $D_{\text{Br}}/D_{\text{Cl}}\simeq0.5$. Note that average value of $D/J$ changes with doping, i.e., realized values $x=0.0, 0.06, 0.08, 0.13$ \cite{Yu2012,Wulf2013,Povarov2015} have $D/J\simeq 4.0,3.3,3.0,2.6$, respectively. For the maximal $x=0.25$ (concentration $x$ has $2x$ of changed bonds) the system will be in the Haldane--like limit, $D\lesssim J$. As a consequence, with increasing doping our mapping becomes less accurate. However, there are may other candidates of $S=1$ materials with reduced dimensionality and larger single--ion anisotropy, e.g., \mbox {CsFeBr$_3$} with $D/J\sim5$ \cite{Dorner1989}, \mbox{Ni(C$_2$H8N$_2$)$_2$Ni(CN)$_4$} with $D/J\sim7$ \cite{Orendac1995}, or \mbox {Sr$_3$NiPtO$_6$} with $D/J\sim9$ \cite{Chattopadhyay2010}. If successfully doped, these could be even better effective realizations of random magnetic fields. Also, any systems with $S>1/2$ and single--ion anisotropy can be investigated in similar manner, as in the case of \mbox{Cs$_2$CoCl$_4$} compound which can be described by $S=3/2$ Hamiltonian with $D/J\sim 10$ \cite{Breunig2013,Breunig2015}. Another intriguing possibility is engineered magnetic atomic structure on surface, where both, large magnetic anisotropy and exchange interactions were demonstrated (for a review see Ref. \onlinecite{Spinelli2015}). 

Regarding the possibility of MBL effects in DTNX, we stress that several works \cite{Pal2010,Luitz2015,Serbyn2015,Bordia2015,Nandkishore2015,Mott1969,Nandkishore2014,Johri2015,Huse2015,Hyatt2016,Steinigeweg2015,Karahalios2009,Znidaric2008,Bardarson2012,Znidaric2016} suggest that MBL regime for $S=1/2$ model \eqref{s12ham} appears for $\delta\wt{h}/J\gtrsim7$ (with $\delta J=0$), which is not reachable with DNTX  having $D/J=4$ and estimated $\delta\wt{h}/J\sim4$ for
assumed $\delta J = 0$. It is further a future theoretical challenge to explore the effects of higher order terms in $J/D$, higher dimensionality (2D and 3D) \footnote{Although we study 1D system the mapping to $S=1/2$ effective model holds also for higher dimensions, provided that the system has large single--ion anisotropy.} and even more importantly the effects of other degrees of freedom in real compounds, e.g., phonons. In particular, since these might prevent localization \cite{deutsch1991,Prelovsek2016a,Johri2015}.

In summary,  we have shown that $S=1$ system with large single--ion anisotropy and quenched randomness essentially realizes a random local magnetic fields in an effective low--energy $S=1/2$ Hamiltonian. This could be tested by exploring the spin or heat transport or alternatively the nonergodic behavior via the persistent imbalance \cite{Schreiber2015,Choi2016} like quantities, e.g., possibly by NMR or $\mu$SR.

\begin{acknowledgments}
This work was supported by the U.S. Department of Energy, Office of Basic Energy Sciences, Materials Science and Engineering Division, European Union program FP7-REGPOT-2012-2013-1 under grant agreement n. 316165 and by Slovenian Research Agency under program P1-0044. We acknowledge helpful and inspiring discussions with X. Zotos, P. Prelov\v{s}ek, M.~Klanj\v{s}ek, and R.~\v{Z}itko.
\end{acknowledgments}


\end{document}